\documentstyle[preprint,prl,aps]{revtex}
\begin{document}
\draft

\title{Quantum Correlated Interstitials
and the Hall Resistivity of the Magnetically
Induced
Wigner Crystal}
\author{Lian Zheng and H.A. Fertig}
\address{Department of Physics and Astronomy, University of Kentucky,
Lexington,
Kentucky 40506-0055}
\date{\today}

\maketitle
\begin{abstract}
We study a trial wavefunction for an interstitial in
a Wigner crystal. We find that
the electron correlations,
ignored in a conventional Hartree-Fock treatment,
dramatically lower the interstitial energy,
especially at fillings close to an
incompressible liquid state.
The correlation between the interstitial electron and the lattice
electrons
at $\nu <1/m$ is introduced by constructing a trial wavefunction
which bears a Jastrow factor of a Laughlin state at $\nu=1/m$.
For fillings close to but just below $\nu=1/m$, we find
that a perfect Wigner crystal becomes unstable against formation
of such interstitials.
It is argued that conduction due to correlated interstitials
in the presence of weak disorder leads to the {\it classical}
Hall  resistivity, as seen experimentally.
\end{abstract}


\narrowtext

In very strong perpendicular magnetic fields
at low temperatures, it is generally believed that
a two-dimensional electron gas should form a crystalline
lattice, in the absence of disorder.
A growing body
of experimental evidence has accumulated in the last five
years that this so-called ``Wigner crystal''\cite{wigner}
(WC) may have been realized in high-quality
heterojunction systems.  Evidence of this has come in the form of rf
data \cite{andrei,paalanen},
transport experiments\cite{goldman},
cyclotron resonance \cite{besson}, and photoluminescence (PL)
experiments\cite{buhmann,clark}.  One of the very interesting
aspects of this system is the competition between the quantum
mechanical correlations that lead to
the fractional quantum Hall
effect (FQHE), which are elegantly described by the Laughlin trial
wavefunctions\cite{laughlin},
and the types of wavefunctions that describe the Wigner
crystal, which is typically described by a Hartree or Hartree-Fock (HF)
approximation as a starting point\cite{maki}.
In particular, transport data\cite{goldman} have suggested
that there is a very sharp transition between the FQHE and the
WC as one varies the magnetic field
in the vicinity of filling fraction $\nu=1/5$.  Since
the transition occurs over an extremely narrow magnetic
range, it is
natural to ask whether there are any fluctuations
in the crystal state that hint at the transition
if one is at filling fractions (for example) near, but
below the transition.  We address this subject in this
Letter.

Our principle motivation for this study is the expectation that
density fluctuations of the WC in the vicinity of a FQHE
filling fraction should be strongly influenced by Laughlin-like
correlations.  As a paradigm of this, consider the WC at filling
fractions close to, but just below $\nu=1/5$.
Among the basic excitations of
the WC, interstitial electrons in particular
create a local charge density that
is raised closer to, or possibly above, 1/5.  We
expect that interstitial energies should be strongly renormalized by
correlations in this situation\cite{c1}.
To model such effects, we consider
the following trial wavefunction for a WC with a single interstitial:
\begin{equation}
\psi={\cal A}\lbrace\prod^N_{i=1}(u-z_i)^m{\rm e}^{-|u|^2/4}\psi_{WC}(z_1,
\dots,z_N)\rbrace
\label{eqpsi1}
\end{equation}

where $\psi_{WC}(z_1,\dots,z_N)$ is the wavefunction
for a perfect WC of $N$ electrons, with electrons localized near
sites $\vec{R}^o_i$, $z_i=x_i-iy_i$ is an electron coordinate in
complex notation,
$u$ is the interstitial coordinate,
$m$ is an integer chosen such that
$1/m$ is larger than the filling fraction $\nu$
(we show below that the wavefunction no longer represents
an interstitial for $\nu > 1/m$), and ${\cal A}$ is
the antisymmetrization operator.
We work here in units of the magnetic length $l_o=(\hbar c/eB)^{1/2}$,
where $B$ is the applied magnetic field.
We note that
if one uses a Hartree approximation for $\psi_{WC}$, the case $m=0$
corresponds to an antisymmetrized Hartree approximation for the
interstitial.  Such approximate forms for wavefunctions of the
WC at low enough fillings have been argued to be quite good,
since one may show that exchange corrections to the WC energy
at low fillings are quite small\cite{maki}.

We have evaluated the energy of the wavefunction
 in Eq. (\ref{eqpsi1}) for a Coulomb potential approximately
for various fillings and choices of $m$.  Fig. \ref{fig1} illustrates these
energies for $m=5$ and $m=7$ as a function of $\nu$.  The energy
$\epsilon$
plotted here is defined as
$\epsilon = E_{N+1}^{i} - E_{N+1}^{WC}$, where $E_{N+1}^i$ is
the energy of the state with $N$ electrons
and 1 interstitial, $E_{N+1}^{WC}$
is the energy of a perfect lattice of $N+1$
electrons.
Thus, $\epsilon$ represents the energy to deform a perfect lattice
of $N+1$ electrons into a lattice with $N$ electrons and
one interstitial, with no change in the net charge of the system.
For simplicity, we have used a square
rather than a triangular lattice for $\psi_{WC}$;
this should have no effect on our basic conclusions.
In the insert of Fig. \ref{fig1}, the energy $\epsilon$
is shown as a function of the strength of the correlation
parameter $m$.
One can see
in the main figure that the energies are extremely low over a large
range of filling fractions, and clearly become negative
for filling fractions just below $\nu=1/m$.  (For the case
$m=5$, the energy is just slightly negative over a
surprisingly large
range of $\nu$.  Our numerical
error unfortunately does not allow us to conclude that the energies are
actually negative except above $\nu \approx 0.17$).
We have found similar results for $\nu=1/9$.

Several important conclusions immediately follow from these
results: (1) A perfect Hartree-Fock WC cannot be stable for
filling fractions very close to $\nu=1/m$ for $m \le 9$.
Our calculations indicate that, at the very least, a
superlattice state of correlated interstitials would
have lower energy than a perfect HF WC at such fillings.
We will argue below that simply including a Jastrow factor in
front of an otherwise perfect HF WC will not lead to such
a large lowering of the energy.  Such a structural
phase transition may be related to PL anomalies seen for $\nu=1/m$,
but at
much lower fillings than those for which the FQHE is found
in transport\cite{buhmann2}.
(2) Since the energies of the interstitial states are so very
low over a large range of filling fractions, these are likely
to contribute significantly to charge transport at fillings
for which the crystal is stable.  Furthermore, we will argue below
that the Hall resistivity in the presence of weak disorder
takes on the {\it classical} value $\rho_{xy}=h/\nu e^2$,
as is found in experiment.\cite{goldman3}
(3) Experimental observation of a vanishing activation energy
as the filling fraction approaches $1/5$ from below
could be understood
in terms of the correlated interstitials, whose energies vanish
near the FQHE filling fractions.
(4) The well-known\cite{goldman2}
``dip'' in $\rho_{xx}$ near $\nu=1/7$, where to
date no FQHE has been observed, might be understood as
being due to correlated interstitials, which have particularly
low energies near this filling.

To understand why the correlated interstitial has such low energy,
it is convenient to use a variation of the ``plasma
analogy'' first exploited by Laughlin in the context of
the FQHE\cite{laughlin}.  Dropping the antisymmetrization\cite{c2}
in Eq. (\ref{eqpsi1}), and using a Hartree approximation for $\psi_{WC}$,
we write the square modulus in the form
$|\psi|^2=e^{-\beta \Phi}$, where, for $\beta=1$,
\begin{equation}
\Phi=-2m\sum^N_{i=1}{\rm ln}(|u-z_i|)+{|u|^2\over2}
+\sum^N_{i=1}{|z_i-R^o_i|^2\over2}
\label{eqphi1}
\end{equation}
where $R^o_{i}=R^o_{ix}-iR^o_{iy}$ are the perfect WC lattice sites
in complex notation.
This represents the energy of a classical two-dimensional Coulomb
particle with charge 1, interacting with a neutralizing background of charge
density $\sigma_b=1/2\pi$, and a lattice of charges $m$
with charge
density $\sigma_l=m/a^2$, where $a$ is the WC lattice constant.
On a coarse scale, the interstitial is thus interacting with
a system of net charge density $\sigma=(1-m\nu)/2\pi $
($\nu$ is the filling in the {\it absence} of the interstitial),
and thus spreads out in the center of the disk to an area of
$1/\sigma$.  We note that precisely at the filling $\nu=1/m$, the
interstitial will ``spread out'' uniformly over the disk, and for
higher fillings the interstitial electron will have its highest
probability at the system edge, essentially being ejected from
the system.  This is the reason the wavefunctions are only
physically reasonable for fillings below $\nu=1/m$ \cite{hfrk1}.
 Fig. \ref{fig2}
illustrates the probability of finding the interstitial
at a particular position along a crystal symmetry axis for
$m=5$ and various values of $\nu$, which is calculated
using Eq.~(\ref{eqpsi2}) below.

We are now in a position to understand why the correlations lower
the energy of interstitial, especially close to $\nu=1/m$.
In essence, the Jastrow factor allows us to take advantage of
quantum fluctuations in the WC.  When there are configurations
in $\psi_{WC}^N$ that have large holes in them, the interstitial
has a high probability of being located there, especially
when $\nu \sim 1/m$, for which the interstitial ``density''
(Fig. \ref{fig2}) samples a large region of the crystal.  Conversely,
when the crystal forms regions of high density, the interstitial
has a low probability of approaching that region.  In this way,
the interstitial electron does an excellent job of minimizing its
energy.  We note that a Jastrow factor
of exponent $m$, involving all electron pairs,
multiplying a perfect WC HF state would probably not
lower the energy per electron so much as the interstitial
electron energy is lowered when $\nu < 1/m.$  This is because
in the former case the electrons
tend to remain closely tied to their lattice sites, and
hence do not ``sample'' enough of the lattice to take great
advantage of the quantum fluctuations.  Thus, the low
energy of the interstitial wavefunction is due as much to
its locally high density as it is to the good correlations
it has with the lattice electrons.

We now consider the Hall
effect due to correlated interstitials in the
presence of weak disorder, and show this
gives the {\it classical} Hall resistivity,
as seen in experiment.\cite{goldman3} We first need to
determine how disorder is likely to affect
our interstitial wavefunction.  The main source of
disorder in heterostructures is thought to be caused by
fluctuations in the neutralizing background charge density,
which typically is set back from the electron gas by several
thousand Angstroms.  Recent classical calculations of the
electron crystal state in the presence of such disorder have
shown that this leads to small variations of the crystal density
over long length scales.\cite{cha}  As seen in Fig. 1, the interstitial
energetically favors regions of higher density.  We can build
correlations into the wavefunctions to optimize the energy by
altering our correlation factor slightly,
to take the form $\prod_{i}(u-z_i)^{m_i}$, with
$m_i$ chosen to be larger for lattice sites in low density
regions, and smaller elsewhere.  One may show that the
interstitial will not be ejected from the system, provided
$<m_i> \le 1/\nu$, where $<\dots>$ denotes an average over lattice
sites.  The higher energy regions may be optimally avoided if one
chooses $<m_i> = 1/\nu$.

The Hall resistivity of interstitials in such states can be estimated
using a mean-field approximation, in which we replace $m_i$ by
$<m_i>$ for all the lattice sites.
This is justified, because if the interstitials are
flowing in a current, they will pass many of the lattice electrons.
Furthermore, the correlation factors turn out to affect the lattice
electrons over very long distances, as will be described below.
The key insight is that
the correlation factor $(u-z_i)^{m_i}$ makes the interstitial appear
as a magnetic flux tube to lattice electron {\it i},
with strength $m_i\phi_0$, where $\phi_0=hc/e$.
%and the magnetic flux
%is directed opposite to the real magnetic field.
Thus, on average, a single correlated interstitial flowing
through the system should have the same effect as a moving
flux tube of strength $<m_i>=1/\nu$.
It is easy to show
using Faraday's law that an electric current $I$
flowing in the $\hat{x}$ direction in which each
carrier has this magnetic flux will generate  a force on the lattice
electrons, which may be described by an effective voltage drop
of magnitude ${{h} \over {\nu e^2}}I$.
%At any finite temperature,
%this would cause the {\it lattice} electrons to creep in the
%transverse ($\hat{y}$) direction.\cite{cha}
%If one wishes to perform an experiment with current
%flowing purely in the $\hat{x}$ direction, then a {\it real}
%electric field $-{{h} \over {\nu e^2}}I \hat{y}$ must be supplied
%to balance this transverse force.
This results in a finite Hall resistivity $\rho_{xy}={{h} \over {\nu e^2}}$,
as observed experimentally.\cite{goldman3}
We note that a similar mechanism leads to the finite Hall resistance
of the recently proposed ``Hall insulator''.\cite{zhang}

We now discuss in some detail how the energy of
the interstitial state is evaluated.  Using Hartree approximation
for $\psi_{WC}$ in Eq.~(\ref{eqpsi1}) and neglecting the
anti-symmetrization, one has
\begin{equation}
|\psi(u,z_1,\dots,z_N)|^2=
{\rm e}^{-|u|^2/2}\prod^N_{i=1}|u-z_i|^{2m}{\rm e}^{-|z_i-R^o_i|^2/2}
\label{eqpsi2}
\end{equation}
where $R^o_i$ forms a square lattice,
which would be the center of mass of each lattice
electron if the Jastrow-type
correlation factor in the wavefunction were absent.
The interstitial
electron density distribution $\rho(u)=\int\prod^N_{i=1}
d^2z_i|\psi(u,z_1,\dots,z_N)|^2
/\int d^2u\prod^N_{i=1}
d^2z_i|\psi(u,z_1,\dots,z_N)|^2$,
shown in Fig.~\ref{fig2}, can
be directly evaluated from the above expression,
since the integrations over the lattice electron coordinates
may be computed analytically. The energy of this
interstitial state $E^i_{N+1}$ consists of two parts:
$E^i_{N+1}=E_1+E_2$, where $E_1$ represents the Coulomb interaction
between the interstitial electron and the lattice electrons,
and $E_2$
represents the pairwise interactions among the lattice electrons.
Numerical evaluation of  $E_1$ can be carried out directly
from Eq.~(\ref{eqpsi2}) \cite{rkz2}
by analytical integration over the $z_i$'s and numerical integration
over $u$.
An exact calculation for $E_2$, on the other hand,
is highly impractical.
The reasons are easily recognized.
To obtain the interaction energy from one pair of electrons,
one needs to compute a 6-dimensional integral over
the coordinates of the interstitial and the pair of the electrons
involved. This 6-dimensional integral has to be performed for
each of the $(1/2)N(N-1)$ pairs for a system of $N$ electrons.
Moreover, one needs to compute
$E_2$ for relatively large systems,
since $E_2$ turns out to depend on the system size \cite{rkz3}.
We thus turn to an approximate evaluation of $E_2$.

Our approximation scheme is based on the observation that the
main effect of
the Jastrow factor in Eq.~(\ref{eqpsi2}) is to shift
the center of mass of each lattice electron wavepacket
away from
$R^o_i$. Beyond this, these wavepackets are also
deformed from circularly symmetric Gaussian distributions.
Such effects can be systematically accounted for if
we express the Coulumb potential in
terms of a multipole expansion. One immediately realizes
that the effect of the center of mass shift
on $E_2$ is much more important than that of the wavepacket
deformation.  This is because the shift of the center of mass
of each lattice electron changes the dipole and higher-order multipole
moments,
while the wavepacket deformation changes only the quadrupole and higher-
order multipole moments.
Since the extension of the electron wavepacket is much smaller
than the inter-particle distance for $\nu\leq1/5$,
a natural approximation for $E_2$ is
to neglect the wavepacket deformation, {\it ie.},  to assume
that an electron at lattice site $i$
consists of a Gaussian wavepacket localized
at the shifted center of mass $<\vec{r}_i>$, where $<\vec{r}_i>$
is calculated numerically from the probability distribution
Eq.~(\ref{eqpsi2}) as
$<\vec{r}_j>=\int d^2u\prod_id^2z_i\vec{r}_j|\psi|^2/
\int d^2u\prod_id^2z_i|\psi|^2$.
 Therefore, one has $E_2\simeq\sum_{i<j}
V(|<\vec{r}_i>-<\vec{r}_j>|)$, where the effective potential
is taken as that of a HF crystal\cite{maki}
\begin{equation}
V(R)={1\over R}+{1\over R^3}+{9\over2R^5}+{75\over2R^7}+\dots\ .
\label{eqvr1}
\end{equation}

The energy of the correlated interstitial state with
$E_1$ calculated from Eq.~(\ref{eqpsi2}) and $E_2$
calculated from Eq.~(\ref{eqvr1}) is shown in Fig.~\ref{fig1}.
This is our primary result. Before concluding the description
for the calculation, we remind the reader that the
numerical uncertainties, namely,
neglecting the anti-symmetrization,
neglecting part of the higher-order multipole contribution to $E_2$,
as well as the numerical errors, are all relatively small.
More important corrections to the precise energies of
the interstitials would result from using a triangular
lattice, and accounting for lattice relaxation in the vicinity of the
interstitial.
However, such improvements should not
alter the qualitative result.

To summarize, we have studied a trial wavefunction for an interstitial in
a Wigner crystal and showed that
the electron correlations,
ignored in a conventional Hartree-Fock treatment,
may be essential in understanding several
properties of a Wigner crystal, especially at fillings close to an
incompressible liquid state.
As a result of the correlation, the state of a Wigner crystal
with interstitials is found to
have very low energy compared to the corresponding
Hartree-Fock interstitial.
In particular, we have found
that for fillings close to but
below $\nu=1/m$, the correlated interstitial state is actually
energetically favored over
a perfect Wigner crystal.
Away from such fillings, we found in the presence of weak disorder
that thermally excited
interstitials carrying a current generate a transverse
electric field, leading
to a finite Hall resistivity, $\rho_{xy}=h/\nu e^2$,
as observed experimentally.

The authors would like to thank
Dr. A.H. MacDonald for helpful discussion. This work
is supported by NSF through Grant No. DMR-9202255.

\begin{figure}
\caption{}
The energy $\epsilon=E^i_{N+1}-E^{WC}_{N+1}$ as functions
of the Landau level filling factor $\nu$ for $m=5$ and $m=7$,
where $E^{i}_{N+1}$ is the energy of the
$N+1$ electron state described by
Eq.~(\ref{eqpsi2}) and $E^{WC}_{N+1}$ is the energy of a
$N+1$ electron Hartree-Fock Wigner crystal state.  The insert shows
the energy $\epsilon$ at $\nu=0.19$ as a function of the exponent
$m$ of the Jastrow factor, {\it ie}., as a function of the
strength of the correlation.
\label{fig1}
\end{figure}
\begin{figure}
\caption{}
The probability distribution for the interstitial electron
in a correlated interstitial state described by Eq.~(\ref{eqpsi2})
with $m=5$ and $\nu=0.19, 0.17,\ 0.1$.
The distance is along the $y=0$ axis in a
Cartesian coordinate reference frame
where the square lattice
$R^o_{ij}$ is expressed as $(i-0.5,j-0.5)$.
The distance is in unit of the lattice constant $a$.
\label{fig2}
\end{figure}

\begin{references}
\bibitem{c1} Among the basic excitations of the WC, the interstitial has
points of higher local density than other types of excitations.
We thus expect it to be more strongly affected by correlations than
other excitations below but near a FQHE filling fraction.
\bibitem{c2}
We have estimated the exchange correction to the energy. Due to
the Jastrow factor, the correction to the energy from
wavefunction overlap between the interstitial
and any of its nearest lattice electrons is in the order of
1 part in $10^6$ for $\nu < 1/5$.
The small effects of exchange between two
lattice electrons have been known from the work of Maki
and Xotos, Ref. \cite{maki}.
\bibitem{rkz2} For a system size of $N\geq 4\times10^4$, the
accuracy of $E_1$ is better than $0.01{\rm e}^2/l_o$.
\bibitem{rkz3} The system size has to be
large enough to eliminate the boundary effects
on $E_2$. In the actual calculation of $E_2$
described in the following paragraphs,
the system size is chosen as $N=(1/5,2/5,3/5,4/5,1)\times 10^6$.
The value of $E_2$ is then extrapolated to $N=\infty$.
\bibitem{hfrk1} It is interesting to speculate that similar
states could be constructed for vacancies for fillings
just above $\nu=1/m$.
\end{references}
\end{document}